\documentclass[final,3p,times,twocolumn]{elsarticle}
\usepackage{float}
\usepackage{amssymb}
\usepackage{color}
\usepackage{lmodern}
\usepackage{amsmath}
\usepackage{graphicx}
\usepackage[none]{hyphenat}

\begin{document}

\begin{frontmatter}



\title{The transverse field XY model on the diamond chain}
\ead[label2]{pimentel@ufpi.edu.br}
\author[label1]{S\'ergio Mikael V. S. Rosa}
\author[label1]{Jos\'e Pimentel de Lima}
\author[label2]{Natanael C. Costa}
\author[label3,label4]{Lindberg  Lima Gon\c{c}alves}
\address[label1]{Departamento de F\'isica Universidade Federal do Piau\'i, Campus Ministro Petr\^onio Portela, 64049-550, Teresina PI, Brazil}
\address[label2]{Instituto de F\'isica, Universidade Federal do Rio de Janeiro, Caixa Postal 68528, 21941-972 Rio de Janeiro RJ, Brazil}
\address[label3]{Departamento de Engenharia Metal\'urgica e de Materiais, Universidade Federal do Cear\'a, 60455-760, Fortaleza CE, Brazil}
\address[label4]{Campus da Universidade Federal do Cear\'a, Russas, Cear\'a Brazil, CEP: 62.900-000 }
\begin{abstract}
We consider the $s=1/2$ transverse field $XY$ model on the frustrated diamond chain, considering anisotropic exchange parameters between nearest neighbour spins.
To this end, we employ three different methodologies: mean-field approximations, and state-of-the-art exact diagonalizations (ED), and density matrix renormalization group (DMRG) simulations.
Within a mean-field theory, the Hamiltonian is fermionized by introducing the Jordan-Wigner transformation, and the interacting (many-body) terms are approximated to single-particle ones by a Hartree-Fock approach.
We analyze the behavior of the induced and spontaneous magnetization as functions of the external field, investigating the magnetic properties at the ground state, and at finite temperatures.
Interestingly, the mean-field results are in reasonable agreement with the ED and DMRG ones, in particular for the distorted chain, or at an intermediate/large spin anisotropy parameter.
As our key results, we present phase diagrams anisotropy $\times$ magnetic field at zero temperature, discussing the emergence of phases and its quantum critical points.
Finally, our analysis at finite temperature provides a range of parameters in which an unusual behavior of the induced magnetization occurs -- with it increasing as a function of temperature.
This work presents a \textit{global} picture of the XY model on the diamond chain, which may be useful to understand features of magnetism in more complex geometries. 
\end{abstract}

\end{frontmatter}
\twocolumn
\section{Introduction}
\label{Int}
Frustrated quantum magnetism is a subject under intense debate in Condensed Matter, in mainly due to a plethora of phases and properties that such kind of systems may exhibit in their ground state \cite{schmid2017, diep2005}.
Thus, over the past decades, a great experimental effort has been done to identify and characterize frustrated compounds, with these novel materials appearing in many different geometries -- both in quasi-1D chains \citep{miyata12021,cheranovskii2021} or quasi-2D lattices \citep{kan2020} \cite{Muqing2021}.
The azurite is the paradigmatic example for the former case, exhibiting weakly coupled quasi-1D diamond chains of strongly interacting localized electrons\cite{harald2011,kikuchi2004,cheng2017,fujihala2017}.
Within this context, special attention has been given to effective Hamiltonians, in particular to low-dimensional spin systems, from which we expect to understand the nature of these compounds.  

In view of this, recent theoretical studies have investigated frustrated quasi-1D chains -- in particular, the diamond one --, which  present complex ground states\,\cite{okamoto2003,richter2014,lisnyi2014,richter2015,torrico2018,verkholyak2011}.
In spite of this, there are still several open issues, e.g., concerning their critical behavior or the effects of spin anisotropic interactions to the ground state which constitutes the main subject of this work. 
It is important noticing that, although isotropic spin models, such as the isotropic Heisenberg or the XX model\,\cite{verkholyak2011}, have been used to explain some features of frustrated systems, the application of external parameters (as pressure or magnetic field) could lead to anisotropies in the spin exchange couplings.
Consequently, it seems more appropriate to investigate the most general case.
Therefore we have considered the quantum ($s=1/2$) anisotropic XY model on the diamond chain, presenting, as our key result, phase diagrams for different parameters, and comparing them with the results in literature.

The paper is organized as follow: In the next section, we present the model and the details our methodologies, as well as the definition of quantities of interest, as the magnetization.
Here we perform and compare three different numerical approaches: mean-field Hartree-Fock (HF) theory, exact diagonalization (ED), and density matrix renormalization group (DMRG). 
The results are presented in Section \ref{results}, where we discuss the behavior of the induced and spontaneous magnetization at zero and finite temperatures.
Our conclusions and final remarks are given in Section \ref{conclusions}.

\section{Model and method}\label{ModMet}
\subsection{The model}
\begin{figure}[t]
 \includegraphics[width=80mm]{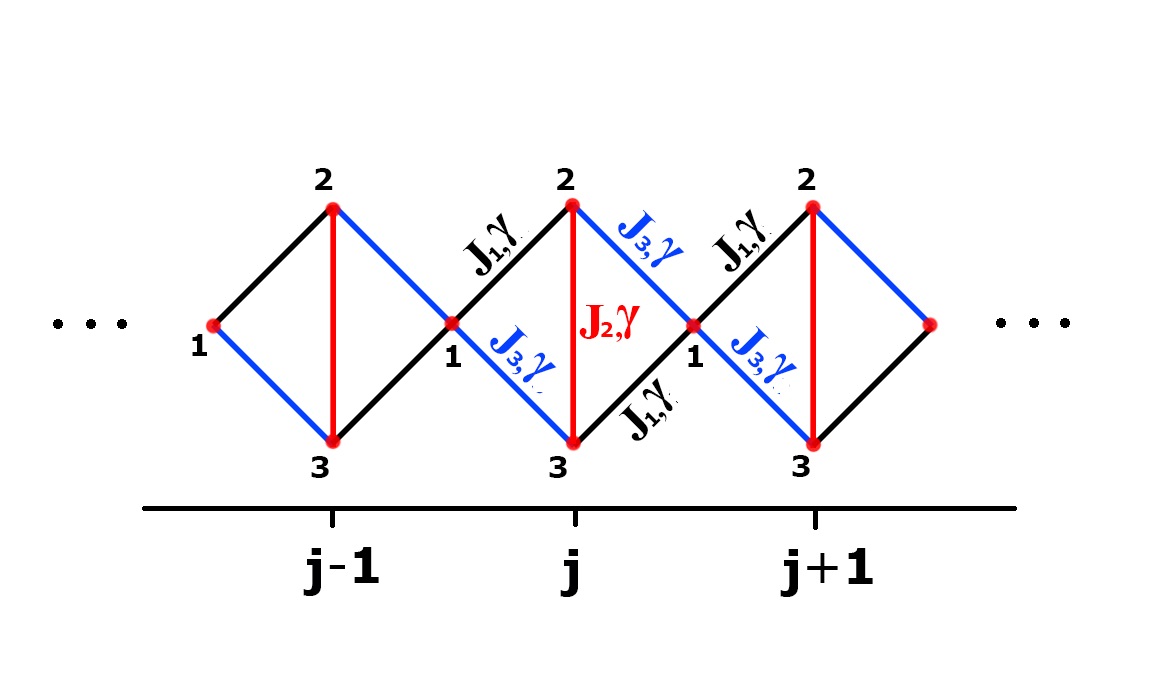}%
 \caption{A schematic representation of $XY$ model $(s=1/2)$ on a diamond chain with three sites $(p=1,2,3)$ per unit cell, $j$.  }
 \label{fig_chain}
 \end{figure}
The diamond chain is a one-dimensional geometry exhibiting unit cells with three basis sites, as displayed in Fig.\,\ref{fig_chain}.
Then, the Hamiltonian of the transverse field $XY$ model on such a chain reads
\begin{eqnarray}
H & =&J_{1}\sum_{j=1}^{N}\big{\{}(1+\gamma)[S_{1,j}^{x}S_{2,j}^{x}+S_{3,j}^{x}S_{1,j+1}^{x}]  \nonumber  \\ 
&+& (1-\gamma)[S_{1,j}^{y}S_{2,j}^{y}+S_{3,j}^{y}S_{1,j+1}^{y}]\big{\}} \nonumber \\ 
&+& J_{2}\sum_{j=1}^{N}\big{\{}(1+\gamma)S_{2,j}^{x}S_{3,j}^{x}+(1-\gamma)S_{2,j}^{y}S_{3,j}^{y}\big{\}} \nonumber \\ 
&+& J_{3}\sum_{j=1}^{N}\big{\{}(1+\gamma)[S_{1,j}^{x}S_{3,j}^{x}+S_{2,j}^{x}S_{1,j+1}^{x}] \nonumber \\ 
&+& (1-\gamma)[S_{1,j}^{y}S_{3,j}^{y}+S_{2,j}^{y}S_{1,j+1}^{y}]\big{\}}+ \nonumber \\ 
&-&h\sum_{j=1}^{N}\sum_{p=1}^{3}S_{p,j}^{z},
\label{eq:spinhamiltonian}
\end{eqnarray}
with $J_{1}$, $J_{2}$ and $J_{3}$ are the exchange couplings between neighbour sites (as shown in Fig.\,\ref{fig_chain}), $\gamma$ the anisotropy parameter, and $h$ is the external transverse field.
$S_{p,j}^{\alpha}$ are standard Heisenberg spin-1/2 operators at a given unit cell $j$ and basis site $p$ ($p$=1,2 or 3), with $\alpha=x,y,z$ denoting the spin components.
Hereafter, we have assume a system with $N$ unit cells under periodic boundary conditions, and define $J_{1}\equiv 1$ as our scale of energy.

By introducing ladder operators in Eq.\,\eqref{eq:spinhamiltonian}, $S_{p,j}^{\pm}=S_{p,j}^{x}\pm iS_{p,l}^{y}$, one may perform a generalized Jordan-Wigner transformation \cite{barbosafilho2001}
\begin{eqnarray}
S_{p,j}^{+}&=&\exp\bigg{(}i\pi \sum_{j'=1}^{j-1}\sum_{p'=1}^{3}c_{p',j'}^{\dag}c_{p',j'} \nonumber \\
&& + \> i\pi \sum_{p'=1}^{p-1}c_{p',j'}^{\dag}c_{ p',j'}\bigg{)}c_{p,j}^{\dag},
\label{eq:jordanwigner}
\end{eqnarray}
in which $c_{p',j'}^{\dag}$ and $c_{p',j'}$ are spinless fermionic creation and annihilation operators, respectively, in the second quantization formalism.
That is, they follow anticommutation relations.
At this point, it is worth mentioning that the sums at the Jordan-Wigner transformation, Eq.\,\eqref{eq:jordanwigner}, follows a given path, which is not unique for chains that have unit cells with more than a single site at the basis, as the diamond one (see, e.g., discussions at Ref.\,\cite{verkholyak2011}).
Here, we adopt the following path: $ \dots \to (p,1) \to (p,2) \to (p,3) \to (p+1,1) \to \dots$, considering the \textit{c}-cyclic case.
Within this transformation, the Hamiltonian becomes $H=H_{1}+H_{2}$, with
\begin{eqnarray}
H_{1} &=& \frac{J_{1}}{2}\sum_{j=1}^{N}\big{[}c_{1,j}^{\dag}c_{2,j}+c_{3,j}^{\dag}c_{1,j+1}   \nonumber \\
&& + \> \gamma(c_{1,j}^{\dag}c_{2,j}^{\dag}+c_{3,j}^{\dag}c_{1,j+1}^{\dag})\big{]} \nonumber \\
&& + \> \frac{J_{2}}{2}\sum_{j=1}^{N}\big{(}c_{2,j}^{\dag}c_{3,j}+\gamma c_{2,j}^{\dag}c_{3,j}\big{)} \nonumber \\
&& + \>  \frac{J_{3}}{2}\sum_{j=1}^{N}\Big{[} c_{1,j}^{\dag}c_{3,j}+c_{2,j}^{\dag}c_{1,j+1} \nonumber \\
&& + \> \gamma \big{(} c_{1,j}^{\dag}c_{3,j}^{\dag}+ c_{2,j}^{\dag}c_{1,j+1}^{\dag}\big{)}\Big{]} + {\rm H.c.} \nonumber \\
&& - \> h \sum_{p=1}^{3}\sum_{j=1}^{N}(n_{p,j}-\frac{1}{2}),
\label{eq:h1}
\end{eqnarray}
and
\begin{eqnarray}
H_{2} &=&-J_{3}\sum_{j=1}^{N}\Big{[} c_{1,j}^{\dag}n_{2,j}c_{3,j} +\> c_{2,j}^{\dag}n_{3,j}c_{1,j+1}\nonumber \\
&& + \>  \gamma \big{(}c_{1,j}^{\dag}n_{2,j}c_{3,j}^{\dag} \nonumber \\
&& + c_{2,j}^{\dag}n_{3,j}c_{1,j+1}^{\dag} \big{)}\Big{]} + {\rm H.c.}.
\label{eq:h2}
\end{eqnarray}
Here \textit{H.c.} denotes the hermitian conjugate, with $n_{p,j}=c_{p,j}^{\dag}c_{p,j}$ being number operators.
As an important remark, one should notice that $H_{1}$ is already in a quadratic form, while $H_{2}$ has the quartic creation/annihilation terms.
From an analytical point of view, $H_{2}$ has the many-particle interaction terms.

Throughout this work we perform three different methods to investigate this system: (1) Mean-field approximations, (2) Lanczos exact diagonalizations, and (3) density matrix renormalization group simulations.
For ED and DMRG, we analyze the ground state of Eq.\,\eqref{eq:spinhamiltonian} at finite size chains, with $N=8$ and 36, respectively.
The DMRG code was implemented and employed by the \textit{ALPS} package\,\cite{ALBUQUERQUE07,Bauer11}.
Concerning the mean-field approach\,\cite{Bruus04}, we performed a Hartree-Fock approximation to deal with $H_{2}$, in the Eq.\eqref{eq:h2}, working at the thermodynamic limit.
It is worth mentioning that, despite being an approximation, recent works have shown that a HF approach could give reliable results in frustrated systems\,\cite{verkholyak2011,sousa17}.
Therefore, we outline the details of our HF methodology in the next subsections.

\subsection{The Hartree-Fock approximation}

First, and foremost, in order to respect the conservation of the number of particles in the limit of $\gamma \to 0$, i.e.~for the limit of the XX model, we need to perform two different HF approaches: (1) a restricted one to terms that do not depend on $\gamma$, and (2) an unrestricted one to those that depend on it.
Therefore, for $c^{\dagger}c^{\dagger}cc$ operators, a restricted HF approximation leads to
\begin{eqnarray}
c_{\alpha}^{\dag} c_{\beta}^{\dag}c_{\beta} c_{\delta}& \approx & - c_{\alpha}^{\dag}c_{\beta} \langle c_{\beta}^{\dag} c_{\delta} \rangle - c_{\beta}^{\dag} c_{\delta} \langle c_{\alpha}^{\dag} c_{\beta} \rangle \nonumber \\
&+& c_{\alpha}^{\dag}c_{\delta}\langle c_{\beta}^{\dag} c_{\beta} \rangle + c_{\beta}^{\dag} c_{\beta} \langle c_{\alpha}^{\dag} c_{\delta} \rangle     \nonumber \\
&+& \langle c_{\beta}^{\dag} c_{\delta} \rangle \langle c_{\alpha}^{\dag} c_{\beta} \rangle - \langle c_{\alpha}^{\dag} c_{\delta} \rangle \langle c_{\beta}^{\dag} c_{\beta} \rangle  ,
\label{Eq:HF1}
\end{eqnarray}
with $\alpha$, $\beta$, and $\delta$ being site indexes $(p,j)$ that appear in the $\gamma$-independent terms on the right-hand side of Eq.\,\eqref{eq:h2}.
Similarly, an unrestricted HF approach to the $c^{\dagger}c^{\dagger}c^{\dagger}c$ (or $c^{\dagger}ccc$) operators that have an explicit dependence on $\gamma$ leads to
\begin{eqnarray}
c_{\alpha}^{\dag} c_{\beta}^{\dag}c_{\beta} c_{\delta}^{\dag}& \approx & - c_{\alpha}^{\dag}c_{\beta} \langle c_{\beta}^{\dag} c_{\delta}^{\dag} \rangle - c_{\beta}^{\dag} c_{\delta}^{\dag} \langle c_{\alpha}^{\dag} c_{\beta} \rangle \nonumber \\
&-& c_{\alpha}^{\dag}c_{\beta}^{\dag}\langle c_{\delta}^{\dag} c_{\beta} \rangle - c_{\delta}^{\dag} c_{\beta} \langle c_{\alpha}^{\dag} c_{\beta}^{\dag} \rangle     \nonumber \\
&+& c_{\alpha}^{\dag}c_{\delta}^{\dag}\langle c_{\beta}^{\dag} c_{\beta} \rangle + c_{\beta}^{\dag} c_{\beta} \langle c_{\alpha}^{\dag} c_{\delta}^{\dag} \rangle     \nonumber \\
&+& \langle c_{\beta}^{\dag} c_{\delta}^{\dag} \rangle \langle c_{\alpha}^{\dag} c_{\beta} \rangle + \langle c_{\alpha}^{\dag} c_{\beta}^{\dag} \rangle \langle c_{\delta}^{\dag} c_{\beta} \rangle  \nonumber \\
&-& \langle c_{\alpha}^{\dag} c_{\delta}^{\dag} \rangle \langle c_{\beta}^{\dag} c_{\beta} \rangle .
\label{Eq:HF2}
\end{eqnarray}

By using Eqs.\,\eqref{Eq:HF1}-\eqref{Eq:HF2}, the quartic $H_{2}$ term can be approximated as
\begin{eqnarray}
H_{2} &\approx&J_{3}\sum_{j=1}^{N}\Big{\{}( A_{1}+\gamma P_{1})(c_{1,j}^{\dag}c_{2,j} +c_{1,j+1}^{\dag}c_{3,j})\nonumber \\
&+&(2A_{2}+2\gamma P_{2})c_{2,j}^{\dag}c_{3,j}+ \nonumber  \\
&-&\bar{n}(c_{1,j}^{\dag}c_{3,j}+c_{1,j+1}^{\dag}c_{2,j}) \nonumber \\
&-&(2A_{3}+2\gamma P_{3})(  c_{2,j}^{\dag}c_{2,j}+c_{3,j}^{\dag}c_{3,j} )  \nonumber \\
&+& \gamma \big{[}A_{1}(c_{1,j}^{\dag}c_{2,j}^{\dag}+ c_{3,j}^{\dag}c_{1,j+1}^{\dag})+2A_{2}c_{2,j}^{\dag}c_{3,j}^{\dag}  \nonumber \\
&-& \bar{n}(c_{1,j}^{\dag}c_{3,j}^{\dag}+c_{2,j}^{\dag}c_{1,j+1}^{\dag})+P_{1} 
\big{]}+H.c.\Big{\}} \nonumber \\
&+& Nu_{0},
\label{eq:h2hartreefock}
\end{eqnarray}
where we have considered that the mean-field parameters are the real part of the averages of EQ.(6) and are given by
\begin{eqnarray}
A_{1}&=&Real(\langle c_{2,j}^{\dag}c_{3,j} \rangle )= Real( \langle c_{3,j}^{\dag}c_{2,j} \rangle), \nonumber \\
A_{2}&=&Real(\langle c_{1,j}^{\dag}c_{2,j} \rangle) =Real(\langle c_{3,j}^{\dag}c_{1,j+1} \rangle), \nonumber \\
A_{3}&=&Real(\langle c_{1,j}^{\dag}c_{3,j} \rangle) =Real(\langle c_{2,j}^{\dag}c_{1,j+1} \rangle), \nonumber \\
\bar{n}&=&\langle c_{2,j}^{\dag}c_{2,j} \rangle=\langle c_{3,j}^{\dag}c_{3,j} \rangle,  \nonumber \\
P_{1}&=&Real(\langle c_{2,j}^{\dag}c_{3,j}^{\dag} \rangle) = Real(\langle c_{3,j}c_{2,j} \rangle), \nonumber \\
P_{2}&=&Real(\langle c_{1,j}^{\dag}c_{2,j}^{\dag} \rangle) = Real(\langle c_{2,j}c_{1,j} \rangle), \nonumber \\
P_{3}&=&Real(\langle c_{1,j}^{\dag}c_{3,j}^{\dag} \rangle) = Real(\langle c_{3,j}c_{1,j} ,\rangle) , \\
\label{parameters}
\end{eqnarray}
and the constant energy term $u_{0}=\frac{3}{2}h-4J_{3}[A_{1}A_{2}-\bar{n}A_{3}] $ $-4\gamma J{3}[A_{1}P_{2} + $
$A_{2}P_{1}-\bar{n}P_{3}]$.
Here, we have imposed that the mean-field Hamiltonian preserves the anisotropic character of model, and therefore we consider the real parameters, as discussed, e.g., by Caux \textit{et. al.}\,\cite{caux2003}.

Introducing a discrete Fourier transform
\begin{eqnarray}
c_{p,j}&=&\frac{1}{\sqrt{N}}\sum_{q}\exp(-\i qj)a_{p,q}, \nonumber \\
a_{p,q}&=&\frac{1}{\sqrt{N}}\sum_{j=1}^{N}\exp(\i qj)c_{p,j}, 
\label{eq:fourier}
\end{eqnarray}
with $q=2\pi j/N$, and $j=0,\pm1,\pm2,\pm3,...,N/2$, our mean-field Hamiltonian becomes
\begin{eqnarray}
H_{HF}&=& \sum_{q}\tilde{H}_{q}  \nonumber \\
&=&\sum_{q} \Big{\{}\big{[}\tilde{J}_{1}+\tilde{J}_{3}e^{iq}\big{]a^{\dag}_{1,q}a_{2,q}}       \nonumber \\
&+& \big{[}\tilde{J}_{1}e^{iq}+\tilde{J}_{3}\big{]}a^{\dag}_{1,q}a_{3,q}+\tilde{J}_{2} a^{\dag}_{2,q}a_{3,q}    \nonumber \\
&+&\gamma\big{[} \tilde{J}_{1}^{\prime}-\tilde{J}_{3}e^{iq} \big{]}a^{\dag}_{1,q}a^{\dag}_{2,-q}        \nonumber \\
&+&\gamma\big{[}\tilde{J}_{3}-\tilde{J}_{1}^{\prime}e^{iq} \big{]}a^{\dag}_{1,q}a^{\dag}_{3,-q}      \nonumber \\
&+&\gamma\tilde{J}_{2} a^{\dag}_{2,q}a^{\dag}_{3,-q} +H.c.\Big{\}}-\sum_{q}\Big{\{}h a^{\dag}_{1,q}a_{1,q}         \nonumber \\
&+&\tilde{h}(a^{\dag}_{2,q}a_{2,q}+a^{\dag}_{3,q}a_{3,q} ) - u_{0}\Big{\}} ,
\label{mf_hamiltonian}
\end{eqnarray}
where we define $\tilde{J_{1}}=\frac{J_{1}}{2}+J_{3}A_{1}+\gamma J_{3}P_{1}$, $\tilde{J_{1}^{\prime}}=\frac{J_{1}}{2}+J_{3}A_{1}$, $\tilde{J_{2}}=\frac{J_{2}}{2}+2J_{3}A_{2}$, $\tilde{J_{3}}=\frac{J_{3}}{2}-J_{3}\bar{n}$, $\tilde{h}=h+2J_{3}A_{3}+2\gamma J_{3}P_{3}$.
To diagonalize the Hamiltonian, we define the transformation~\cite{lieb1961,lima2007}
\begin{eqnarray}
a_{p,q}=\sum_{k} ( \frac{ \psi _{q,k,p}^{*}+ \phi _{q,k,p}^{*}}{2}\eta _{q,k}+  \nonumber \\
\frac{ \psi_{q,k,p}^{*}- \phi_{q,k,p}^{*}}{2}\eta_{-q,k}^{\dag}), \\
a_{p,q}^{\dag}=\sum_{k} ( \frac{\psi_{q,k,p}+\phi_{q,k,p}}{2}\eta_{q,k}^{\dag}+  \nonumber \\
\frac{\psi_{q,k,p}-\phi_{q,k,p}}{2}\eta_{-q,k}),
\label{tcanonical}
\end{eqnarray}
where $\eta_{q,k}^{\dag} $ and $\eta_{q,k} $ are also fermionic creation and annihilation operators, respectively, that leads to $\tilde{H}_{q}=\sum_{k} E_{q,k} \eta_{q,k}^{\dag}\eta_{q,k}+constant$.
One may show that $\phi_{q,k,p}$ and $\psi_{q,k,p}$ are components of the eigenvectors $\mathbf{\Phi}_{q,k}$ and $\mathbf{\Psi}_{q,k}$, from which
\begin{eqnarray}
(\mathbf{A}_{q}-\mathbf{B}_{q})(\mathbf{A}_{q}+\mathbf{B}_{q})\mathbf{\Phi}_{q,k}=E_{q,k}^{2}\mathbf{\Phi}_{q,k}, \nonumber \\
(\mathbf{A}_{q}+\mathbf{B}_{q})(\mathbf{A}_{q}-\mathbf{B}_{q})\mathbf{\Psi}_{q,k}=E_{q,k}^{2}\mathbf{\Psi}_{q,k}, 
\label{eigen}
\end{eqnarray}
with $E_{q,k}^{2}$ being the eigenvalues of the Hamiltonian $\tilde{H}_{q}$, while $\mathbf{A}_{q}$ and $\mathbf{B}_{q}$ being the matrices
\begin{eqnarray}
\mathbf{A}_{q} =   \nonumber \\
\begin{bmatrix}
-h & \tilde{J}_{1}+\tilde{J}_{3}e^{iq} & \tilde{J}_{1}e^{iq}+\tilde{J}_{3} \\
\tilde{J}_{1}+\tilde{J}_{3}e^{-iq} & -\tilde{h} & \tilde{J}_{2} \\
\tilde{J}_{1}e^{-iq}+\tilde{J}_{3} & \tilde{J}_{2} &- \tilde{h} 
\end{bmatrix},
\end{eqnarray}
and
\begin{eqnarray}
\mathbf{B}_{q} =   \nonumber  \\
\gamma
\begin{bmatrix}
0 & \tilde{J}_{1}^{'}-\tilde{J}_{3}e^{iq} & -\tilde{J}_{3}+\tilde{J}_{1}^{'}e^{iq} \\ 
-\tilde{J}_{1}^{'}+\tilde{J}_{3}e^{-iq} & 0 & \tilde{J}_{2} \\
\tilde{J}_{3}-\tilde{J}_{1}^{'}e^{-iq} &- \tilde{J}_{2} & 0 
\end{bmatrix}.
\end{eqnarray}
One may also show that the eigenvectors $\mathbf{\Phi}_{q,k}$ and $\mathbf{\Psi}_{q,k}$ are related by $(\mathbf{A}_{q}+\mathbf{B}_{q})\mathbf{\Psi}_{q,k}=E_{q,k}\mathbf{\Phi}_{q,k}$ \cite{lima2007}.

\subsection{Self-consistent calculation}

At this point, one should notice that the mean-field parameters defined in Eq.\,\eqref{parameters} should be obtained self-consistently.
To this end, we have to obtain the real-space creation/annihilation operators as linear combinations of $\eta$ and $\eta^{\dagger}$, which are
\begin{eqnarray}
c_{p,j}^{\dag}&=&\frac{1}{\sqrt{N}}\sum_{q,k}e^{iqj}( \frac{\psi_{q,k,p}+\phi_{q,k,p}}{2}\eta_{q,k}^{\dag}  \nonumber \\
&+& \frac{\psi_{q,k,p}-\phi_{q,k,p}}{2}\eta_{-q,k}), \nonumber \\
c_{p,j}&=&\frac{1}{\sqrt{N}}\sum_{q,k}e^{-iqj}( \frac{ \psi _{q,k,p}^{*}+ \phi _{q,k,p}^{*}}{2}\eta _{q,k}  \nonumber \\
&+&\frac{ \psi_{q,k,p}^{*}- \phi_{q,k,p}^{*}}{2}\eta_{-q,k}^{\dag}). 
\label{eq:AeB}
\end{eqnarray}
Taking this into account, and using the Fermi-Dirac distribution, the mean-field parameters become
\begin{eqnarray}
A_{1}&=&-\frac{1}{4N}\sum_{q,k}f_{q,k}^{+}(3,2)\tanh\bigg(\frac{\beta E_{q,k}}{2}\bigg) ,\nonumber \\
A_{2}&=&-\frac{1}{4N}\sum_{q,k}f_{q,k}^{+}(2,1))\tanh\bigg(\frac{\beta E_{q,k}}{2}\bigg), \nonumber \\
A_{3}&=&-\frac{1}{4N}\sum_{q,k}f_{q,k}^{+}(3,1)\tanh\bigg(\frac{\beta E_{q,k}}{2}\bigg), \nonumber \\
\bar{n}&=&-\frac{1}{2N}\sum_{q,k}\psi_{q,k,2}\phi_{q,k,2}^{*} \tanh\bigg(\frac{\beta E_{q,k}}{2}\bigg) + \frac{1}{2},             \nonumber \\
P_{1}&=&-\frac{1}{4N}\sum_{q,k}f_{q,k}^{-}(3,2)\tanh\bigg(\frac{\beta E_{q,k}}{2}\bigg), \nonumber \\
P_{2}&=&-\frac{1}{4N}\sum_{q,k}f_{q,k}^{-}(2,1))\tanh\bigg(\frac{\beta E_{q,k}}{2}\bigg), \nonumber \\
P_{3}&=&-\frac{1}{4N}\sum_{q,k}f_{q,k}^{-}(3,1)\tanh\bigg(\frac{\beta E_{q,k}}{2}\bigg),
\label{autoconsistente}
\end{eqnarray}
where $\beta=\frac{1}{k_{B}T}$ (with $k_{B} \equiv 1$) is the inverse of the temperature $T$, and
\begin{eqnarray}
f_{q,k}^{\pm}(p_{1},p_{2})=\psi_{q,k,p_{1}}\phi_{q,k,p_{2}}^{*}\pm\psi_{q,k,p_{2}}\phi_{q,k,p_{1}}^{*}.
\end{eqnarray}
In order to obtain Eq.\,\eqref{autoconsistente}, we have defined quantities in Eq.\eqref{parameters} in their Hermitian form, as real numbers.
For instance, one may define $A_{1}=\langle c_{2,j}^{\dag}c_{3,j} + c_{3,j}^{\dag}c_{2,j}  \rangle/2$, and $P_{1}=\langle c_{1,j}^{\dag}c_{3,j}^{\dag}+c_{3,j}c_{1,j} \rangle/2$.
Equation \eqref{autoconsistente} provide us a set of nonlinear coupled equations, which are solved numerically, therefore, determining the mean-field quantities.

\subsection{Induced and spontaneous magnetizations}
\label{inducedmag}

The magnetic properties are investigated by means of the induced and spontaneous magnetizations, as well as by their respective susceptibilities.
The induced magnetization per unit cell is defined as
\begin{align}
M^{z}=\frac{1}{3N}\sum_{j=1}^{N}\sum_{p=1}^{3}\langle M_{p,j}^{z}\rangle ,
\end{align}
with
\begin{align}
M_{p,j}^{z}=\langle S_{p,j}^{z}\rangle=-\frac{1}{2}\sum_{q,k}\psi_{q,k,p}\phi_{q,k,p}^{*}(1-2n_{q,k}),
\label{mz_average}
\end{align}
and $n_{q,k}=\frac{1}{e^{\beta E_{q,k}}+1}$ being the fermion occupation number.
It leads to
\begin{align}
M^{z}=-\frac{1}{6N}\sum_{q,k,p}\psi_{q,k,p}\phi_{q,k,p}^{*}\tanh(\frac{\beta E_{q,k}}{2}).
\label{av_cell_mag}
\end{align}
Given this, the isothermal susceptibility $\chi_{T}^{z}$ can be obtained by
\begin{align}
\chi_{T}^{z}=\frac{\partial M^{z}}{\partial h},
\label{isoth_susc}
\end{align}
from which we identify the critical points by their singularities.

The analysis of the spontaneous magnetization is more subtle.
We recall that, when we make the transformation $S_{p,j}^{x} \rightarrow -S_{p,j}^{x}$, the Hamiltonian of the Eq.\,\eqref{eq:spinhamiltonian} remains invariant, which leads to $\langle S_{p,j}^{x}\rangle=0$.
Then, to obtain $M^{x}$, we need to calculate the static correlation functions $\langle S_{p,j}^{x}S_{p,j+r}^{x}\rangle$ in the limit $r\rightarrow\infty$.
That is, we define the spontaneous magnetization per cell as
\begin{eqnarray}
M^{x}=\frac{1}{3}\sum_{p=1}^{3} M_{p}^{x},
\end{eqnarray}
where
\begin{eqnarray}
M_{p}^{x}=\lim_{r \to \infty} M_{p}^{x}(r)= \lim_{r \to \infty}  \sqrt{\langle S_{p,j}^{x}S_{p,j+r}^{x}\rangle}.
\end{eqnarray}

To this end, we should use Eq.\,\eqref{eq:jordanwigner} and Eq.\,\eqref{eq:AeB} to define
\begin{eqnarray}
A_{p,j}= c_{p,j}^{\dag}+c_{p,j}  \nonumber    \\
B_{p,j}= c_{p,j}^{\dag}-c_{p,j} ,
\label{AeB2}
\end{eqnarray}
and, consequently, to write
\begin{eqnarray}
\langle S_{p,j}^{x}S_{p,j+r}^{x}\rangle=\frac{1}{4}\langle B_{p,j}A_{p+1,j}B_{p+1,j}B_{p+2,j} \nonumber \\
\times B_{p+2,j} \cdots A_{p-1,j+r}B_{p-1,j+r}A_{p,j+r}\rangle .
\label{correlr}
\end{eqnarray} 
Following Ref.\,\cite{lieb1961}, by applying the Wick theorem, the previous expression becomes
\begin{eqnarray}
[M_{p}^{x}(r)]^{2}=\langle S_{p,j}^{x}S_{p,j+r}^{x}\rangle=   \nonumber  \\
\frac{1}{4}det
\begin{bmatrix},
G(1,1) &G(1,2) &\cdots &   G(1,3r)\rangle\\ 
G(2,1) &G(2,2) &\cdots &   G(2,3r)\rangle\\ 
\vdots & \vdots &\ddots &  \vdots\\ 
G(3r,1) &G(3r,2) &\cdots &   G(3r,3r)\rangle\\ 
\end{bmatrix},
\label{eq:det}
\end{eqnarray}
where $G(l,m)$ are contractions given by $\langle B_{p_{1},j_{1}}A_{p_{2},j_{2}}\rangle$, with $j_{1}=\lfloor(l+p+1)/3\rfloor-1 $, $j_{2}=\lfloor(m+p+2)/3\rfloor-1 $, $p_{1}=l+p-3j_{1}-1$, and $p_{2}=m+p-3j_{2}$.
Here, the symbol $\lfloor \mathcal{\dots} \rfloor$ represents the integer part of a given non-negative quantity, while the contractions are given by
\begin{eqnarray}
\langle B_{p_{1},j_{1}}A_{p_{2},j_{2}}\rangle=\frac{1}{N}\sum_{q,k}e^{iq(j_{1}-j_{2})}\times \nonumber \\
\phi_{q,k,p_{1}}\psi{q,k,p_{2}}^{*}(2n_{q,k}-1).
\label{contractions}
\end{eqnarray}

\section{The results}
\label{results}

\begin{figure}[t]
 \includegraphics[width=85mm]{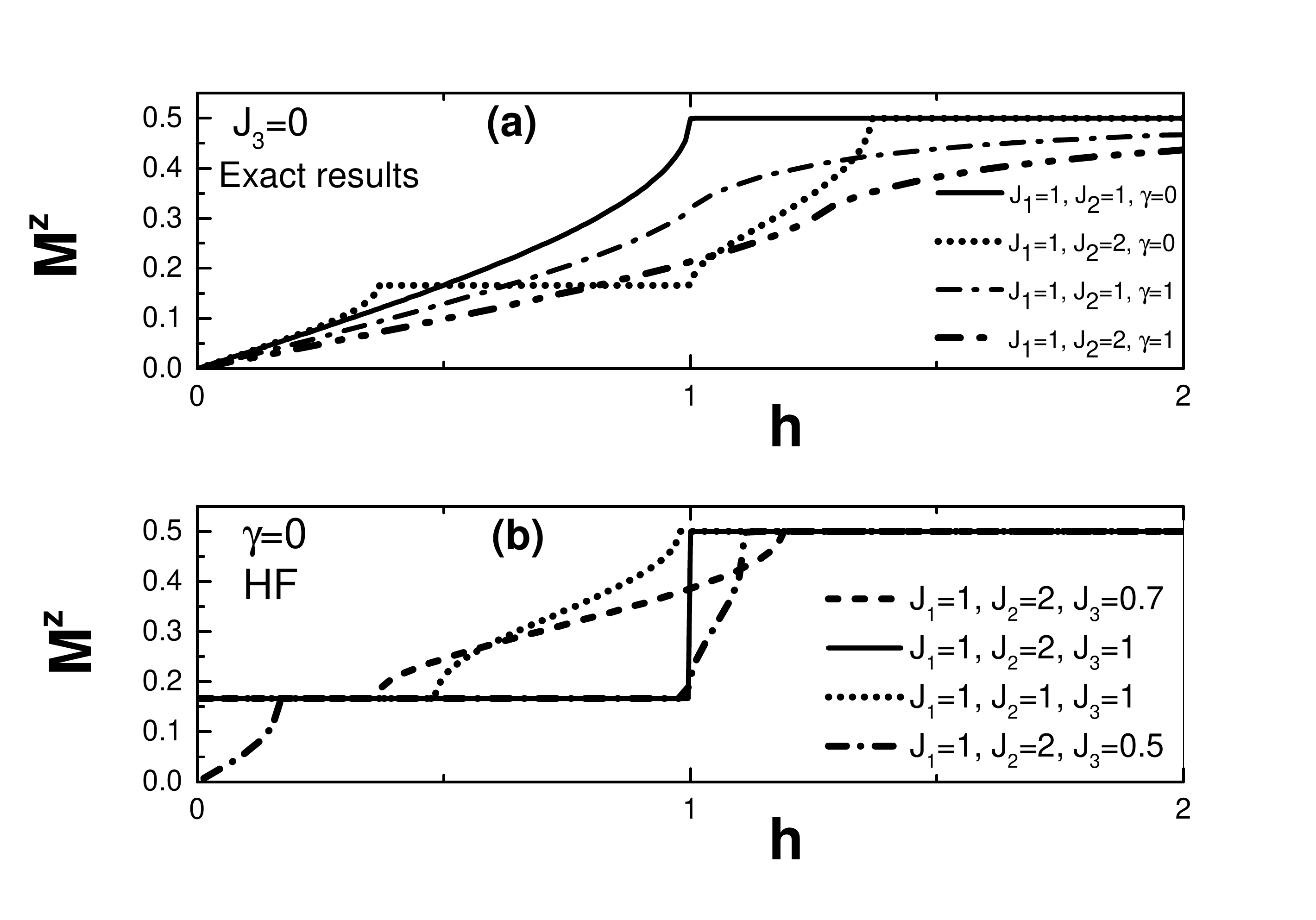}%
 \caption{\label{fig_limit_cases} The induced magnetization $M^{z}$ as a function of the field $h$ for limit cases of the Hamiltonian of the Eq. (\ref{eq:spinhamiltonian}).\textbf{(a)}$J_{3}=0$ for all cases and the model was exactly solved; \textit{continuous line:}$J_{1}=J_{2}=1$ and  $\gamma=0$ is the homogeneous $1D$ $XX$ model; \textit{dash dot line:} $J_{1}=J_{2}=1$ and  $\gamma=1$ is the homogeneous $1D$ $XY$ model; \textit{short dot line:} $J_{1}=J_{2}=2$ and  $\gamma=0$ is the periodic $XX$ model on a chain with three sites per unit cell; \textit{dash dot dot line:} $J_{1}=J_{2}=2$ and  $\gamma=1$ is the periodic $XY$ model on a chain with three sites per unit cell.
\textbf{(b)} $\gamma=0$, that is, the $XX$ model on a diamond chain, in the all cases and the results was obtained by Hartree-Fock approximation \cite{verkholyak2011}; \textit{dash line:} $J_{1}=J_{2}=J_{3}=1$ is a symmetrical diamond chain;   \textit{dot line:} $J_{1}=1$,$J_{2}=2$, $J_{3}=1$ is a symmetrical diamond chain; \textit{continuous line:} $J_{1}=1$,$J_{2}=2$, $J_{3}=0.7$ is a distorted diamond chain; \textit{dash dot line:} $J_{1}=1$,$J_{2}=2$, $J_{3}=0.5$ is a distorted diamond chain.}
 \end{figure}

First, we examine properties at the ground state, starting with limit cases of the model.
Notice that, for $J_{3}=0$ in the Eq.\eqref{eq:spinhamiltonian}, the diamond chain turns into a linear periodic one (with period three), which was solved exactly by two of the authors in Ref.\,\cite{lima2007}, for any values of $J_{1}, J_{2}$, and $\gamma$.
Then, by performing the same procedure outline in Ref.\,\cite{lima2007}, we present in Fig.\,\ref{fig_limit_cases}\,(a) exact results for the diamond chain with fixed $J_{3}=0$, exhibiting the behavior of the induced magnetization as a function of the external transverse field $h$.
For instance, fixing $J_{1}=1$, $J_{2}=2$, and $\gamma=0$ (dotted line), the system starts at a gapless phase for weak $h$, i.e. $\chi_{T}^{z} \neq 0$, until it reaches a gapped 1/3-plateau phase at $ 0.3660... \leq h \leq 1.0 $ \cite{lima2006}.
Further increase of the external field leads again to a gapless phase, until it completely saturates at $h=1.3660...$.
By contrast, for $\gamma=1$, the 1/3-plateau phase disappears, and the saturated phase is obtained only asymptotically.
These results point out the main role of the anisotropy term $\gamma$: smearing plateaus at $M^{z}$. 

For $J_{3} \neq 0$, no exact (analytical) solution is known at the present time, which requires the analysis by numerical methods.
Interestingly, when keeping $J_{3} \ll 1$, the HF results are in very good agreement with those from ED and DMRG (not shown).
It enables us to use such an approximation to investigate the properties of the model in this range of parameters.
Indeed, it is expected that mean-field methods could provide reliable results when the strength of the many-body interacting term is weak.

At this point, we recall that, in the limit of $\gamma=0$ -- and for any set of $J_{\alpha}$--, the Hamiltonian of the Eq.\eqref{eq:spinhamiltonian} becomes the isotropic one, i.e.~it is the $XX$ model.
As discussed in Ref.\,\cite{verkholyak2011}, the HF approximation for this case provides good results only for the distorted chain ($J_{1}\neq J_{3}$), with the symmetrical case ($J_{1}=J_{3}$) being a challenge.
As displayed in Fig.\,\ref{fig_limit_cases}\,(b), the induced magnetization exhibits an 1/3-plateau in absence of external field ($h=0)$ as the ratio $J_{3}/J_{1}\to 1$.
Such a behavior is clearly an artifact of the mean-field approximation.

 \begin{figure}[t]
 \includegraphics[width=80mm]{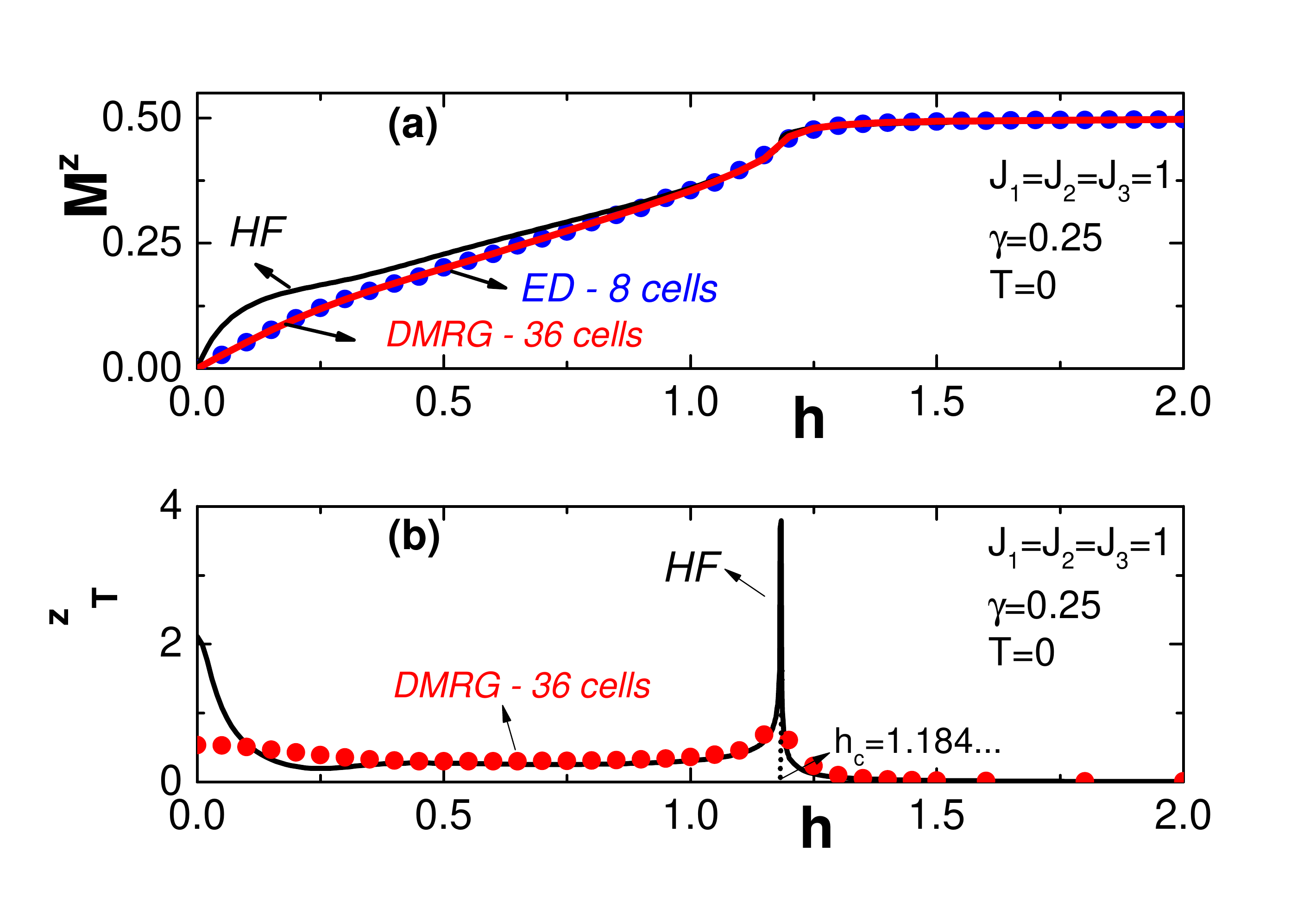}%
 \caption{\label{fig_mz_susc_j1_1_j2_1_j3_1_g_025} Results for symmetrical diamond chain with $J_{1}=J_{2}=J_{3}=1$, $\gamma=0.25$, at $T=0$, \textbf{(a)} the induced magnetization $M^{z}$ as a function of the field $h$, \textit{black line}: calculated by Hartree-Fock approximation (HF) from the Eq.(\ref{mz_average});  \textit{red line}: result from the DMRG for a chain with $36$ cells; \textit{blue points}: result from the exact diagonalization (ED) for a closed chain with $8$ cells  \textbf{(b)} the isothermal susceptibility as a function of the field $h$; \textit{black line:} result of Hartree-Fock approximation (HF); \textit{red points:} result from the DMRG for a chain with $36$ cells.}
 \end{figure}

Given this, it is worth to continue our analysis of the XY model for the most challeging case: the symmetrical one.
Figure \ref{fig_mz_susc_j1_1_j2_1_j3_1_g_025}\,(a) presents the results for $M^{z}$ as a function of $h$, at fixed $J_{1}=J_{2}=J_{3}=1$, and $\gamma=0.25$.
For this intermediate value of $\gamma$, the HF solution (black solid line) does not provide any evidence of the 1/3-plateau, and exhibits just a single critical point at $h_{c}\approx 1.184$, from a gapless phase to the saturated one, as displayed in Fig.\,\ref{fig_mz_susc_j1_1_j2_1_j3_1_g_025}\,(b).
In order to verify it, we have performed ED and DMRG simulations for finite size systems with $N=8$ and $36$ unit cells, respectively.
As shown in Fig.\,\ref{fig_mz_susc_j1_1_j2_1_j3_1_g_025}\,(a) both ED (blue solid circles) and DMRG (red solid line) results agree with those from the HF approach, i.e.~there is such an 1/3-plateau phase.
In addition, as presented in the Fig.\,\ref{fig_mz_susc_j1_1_j2_1_j3_1_g_025}\,(b), the signatures of the HF and DMRG critical points also agree with each other.
When $\gamma$ is increased, the agreement between the mean-field and the unbiased methodologies is improved (not shown). 
These results suggest that, by contrast with the isotropic case, the HF solution for the XY model gives reliable results for the induced magnetization, even for the most challeging case (with $J_{1}=J_{2}=J_{3}$).
 
 \begin{figure}[t]
 \includegraphics[width=85mm]{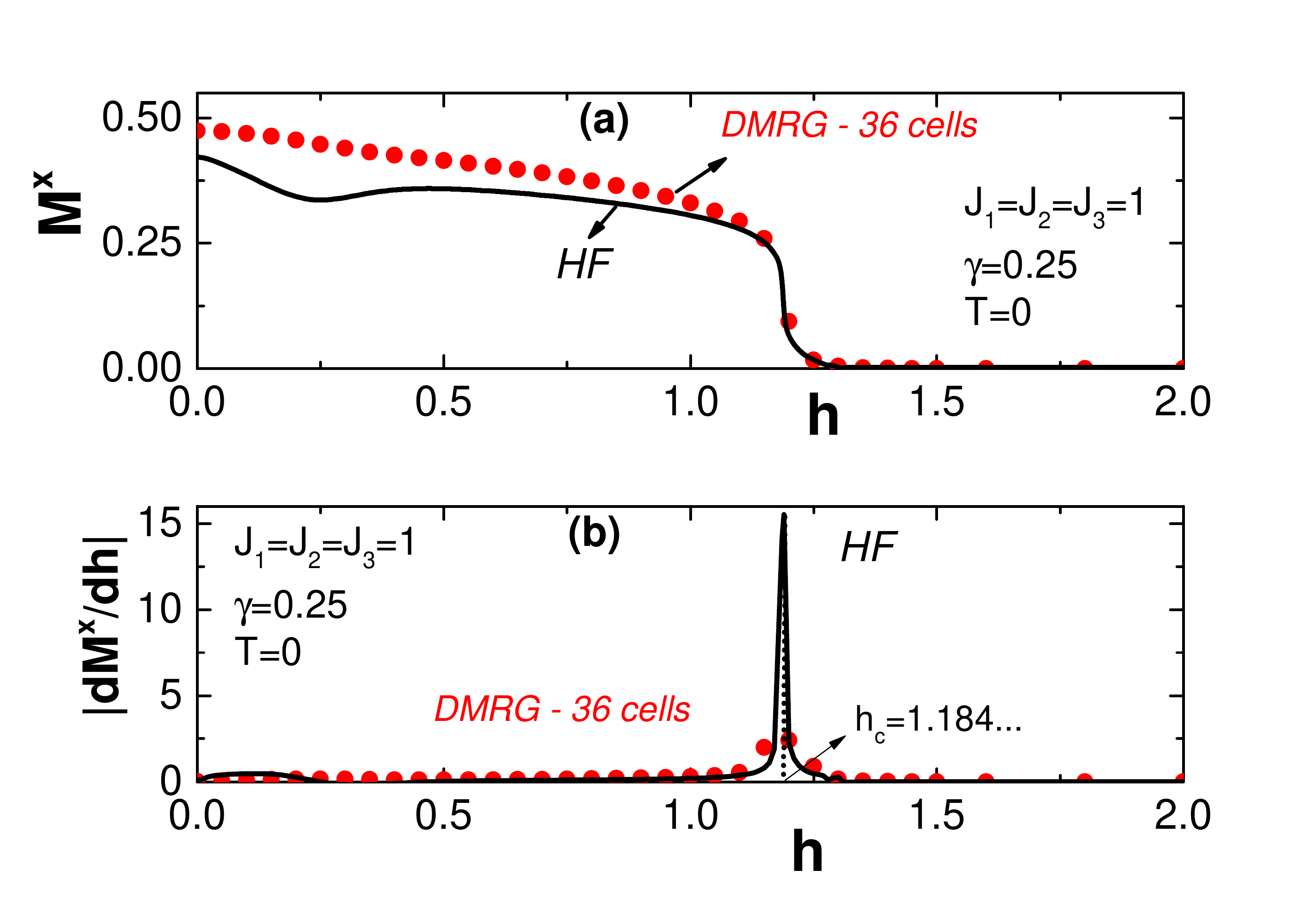}%
 \caption{\label{fig_mx_susc_j1_1_j2_1_j3_1_g_025} Results for symmetrical  diamond chain with $J_{1}=J_{2}=J_{3}=1$, $\gamma=0.25$, at $T=0$, \textbf{(a)} the spontaneous magnetization $M^{x}$ as a function of the field $h$, \textit{black line}: calculated by Hartree-Fock approximation (HF) from the Eqs.(\ref{eq:det} and \ref{correlr}) for $r=30$ cells;  \textit{red points}: result from the DMRG calculus for a chain with $36$ cells;  \textbf{(b)} $|dM^{x}/dh|$ as function of the field $h$, where \textit{black line:} result of the Hartree-Fock approximation (HF); \textit{red points:} result from the DMRG for a chain with $36$ cells.}
 \end{figure}

Proceeding, it is well known that, unlike the XX model, the XY model exhibits spontaneous magnetization.
Then, it is also important to examine the behavior of $M^{x}$, in particular when investigating the critical points.
Figure \ref{fig_mx_susc_j1_1_j2_1_j3_1_g_025} presents our results for (a) the spontaneous magnetization $M^{x}(r)$, and (b) its susceptibility, $|dM^{x}/dh|$, as a function of $h$.

As in the analysis of $M^{z}$, we obtain a reasonable agreement between HF, ED and DMRG for the $M^{x}$ behavior, and for the occurence a critical point.

 \begin{figure}[t]
 \includegraphics[width=85mm]{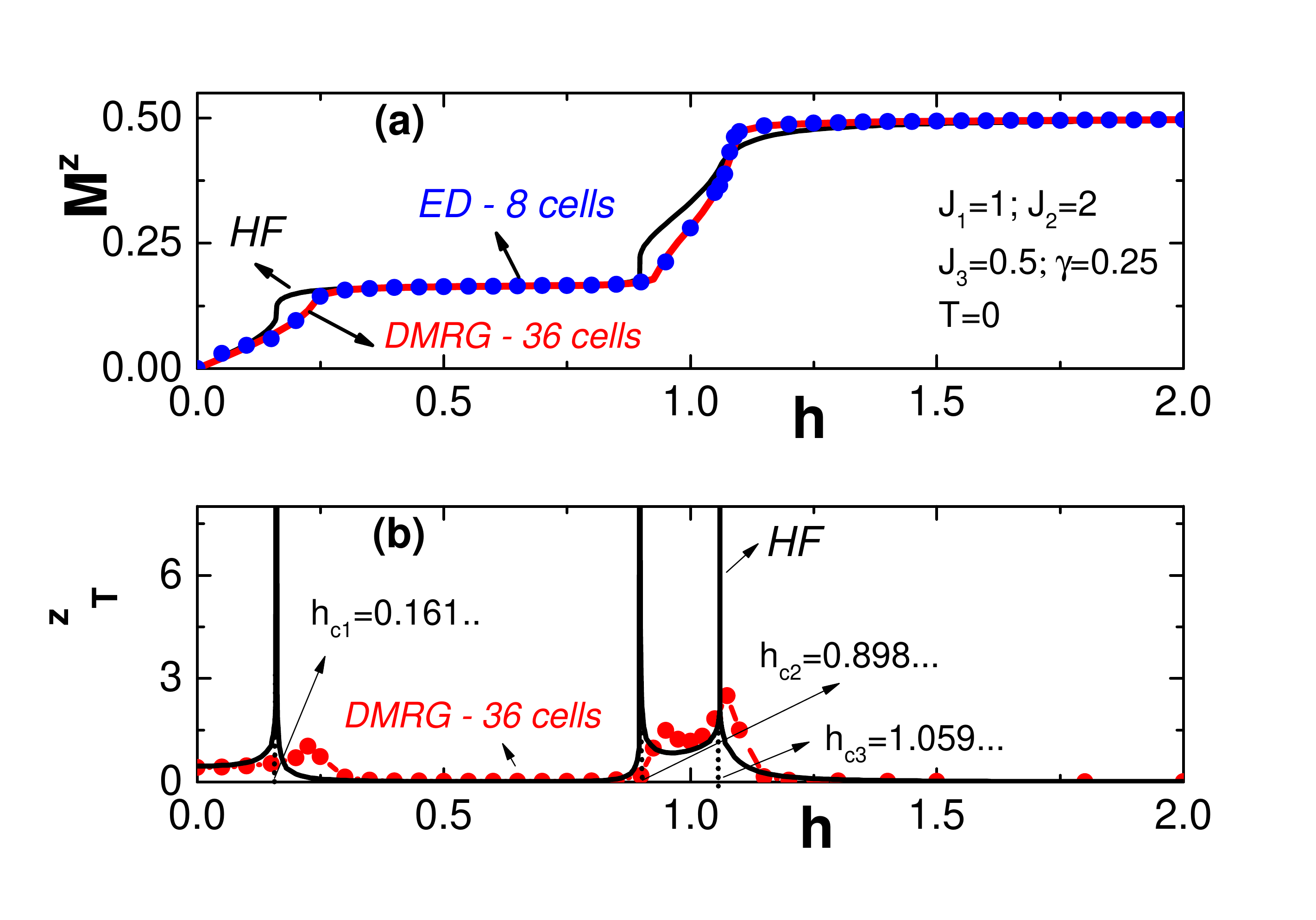}%
 \caption{\label{fig_mz_susc_j1_1_j2_2_j3_05_g_025} Results for a distorted  diamond chain with $J_{1}=J_{2}=J_{3}=1$, $\gamma=0.25$, at $T=0$, \textbf{(a)} the induced magnetization $M^{z}$ as a function of the field $h$, \textit{black line}: calculated by Hartree-Fock approximation (HF) from the Eq.(\ref{mz_average});  \textit{red line}: result from the DMRG for a chain with $36$ cells; \textit{blue points}: result from the exact diagonalization (ED) for a closed chain with $8$ cells  \textbf{(b)} the isothermal susceptibility as a function of the field $h$, where \textit{black line:} result of Hartree-Fock approximation (HF); \textit{red points:} result from the DMRG for a chain with $36$ cells.}
 \end{figure}

We now turn to discuss the distorted case, by setting $J_{1}=1$, $J_{2}=2$, $J_{3}=0.5$ and $\gamma=0.25$.
Figure \ref{fig_mz_susc_j1_1_j2_2_j3_05_g_025} displays (a) the induced magnetization $M^{z}$, and (b) its susceptibility as a function of $h$.
Comparing these results with those from the Fig.\,\ref{fig_limit_cases}, we see that the influence of plateau behaviour of $M^{z}$ is still strong~\cite{lima2006}.
As shown in panel (a), the 1/3-plateau for $M^{z}$ is stronger than the symmetrical case, with both methodologies agreeing with the existence of this phase.
In line with this, the analysis of $\chi_{T}^{z}$ in panel (b) provides three critical points -- $h_{c_{1}}\approx 0.161$, $h_{c_{2}}\approx 0.898$, and $h_{c_{3}}=1.059$, for the HF case --, from which we may identify the phases (gapless, gapped, or saturated) at the ground state.
Despite the quantitative slight difference between HF and DMRG for the values of critical points, it is remarkable to notice that the former exhibits reliable results.   

These critical points are also confirmed by the analysis of the spontaneous magnetization, as displayed in Fig.\,\ref{fig_mz_susc_j1_1_j2_2_j3_05_g_025} for (a) the spontaneous magnetization $M^{x}(r)$, and (b) $|dM^{x}/dh|$ as a function of the field $h$.
From these results, one may see clearly that the magnetization, and its critical points are slightly shifted to the left for the DMRG data.

   \begin{figure}[t]
 \includegraphics[width=85mm]{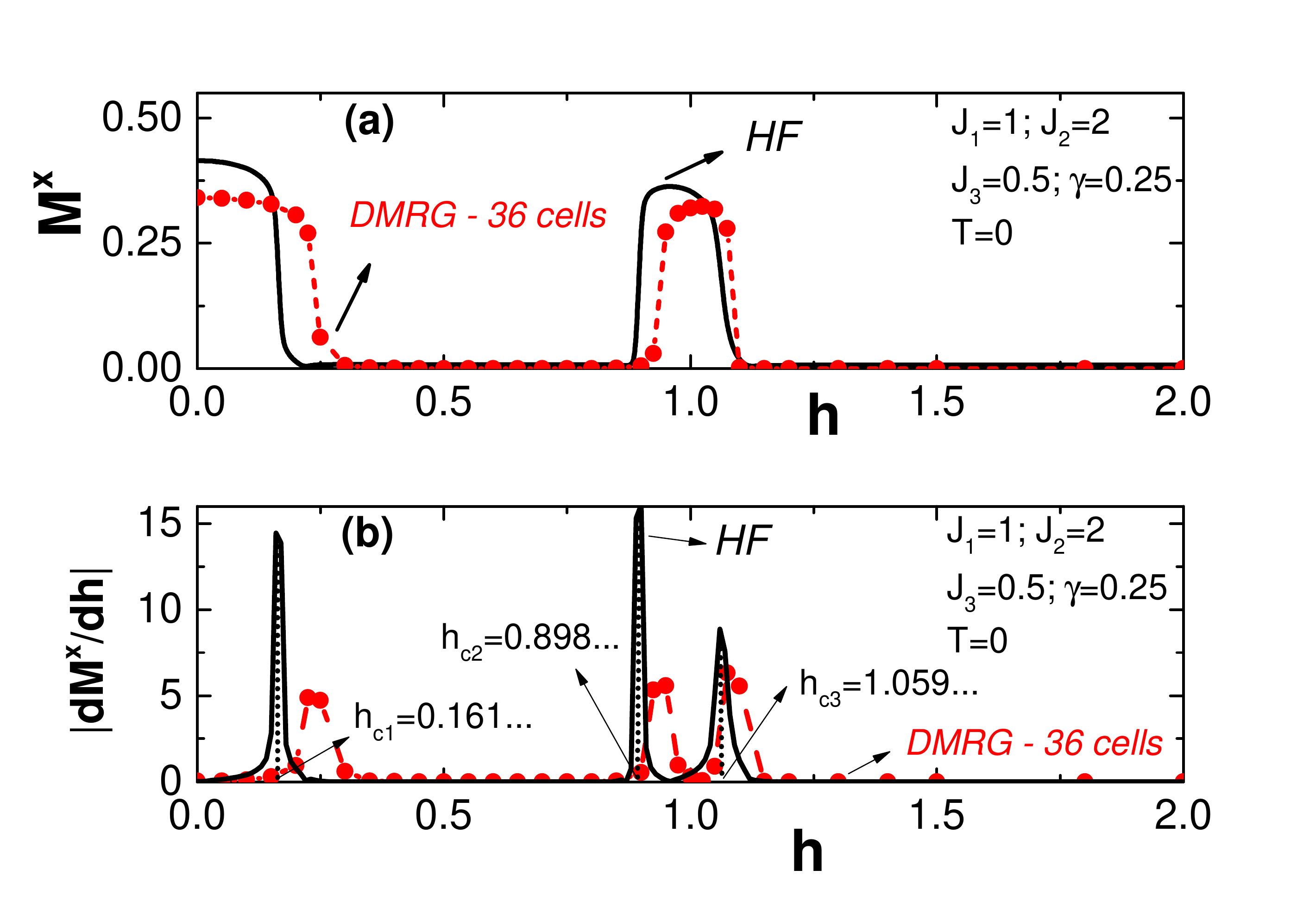}%
 \caption{\label{fig_mx_susc_j1_1_j2_2_j3_05_g_025}Results for distorted  diamond chain with $J_{1}=1$, $J_{2}=2$, $J_{3}=0.5$, $\gamma=0.25$, at $T=0$, \textbf{(a)} the spontaneous magnetization $M^{x}$ as a function of the field $h$, \textit{black line}: calculated by Hartree-Fock approximation (HF) from the Eqs.(\ref{eq:det} and \ref{correlr}) for $r=30$ cells;  \textit{red points}: result from the DMRG calculus for a chain with $36$ cells;  \textbf{(b)} $|dM^{x}/dh|$ as function of the field $h$, where \textit{black line:} result of the Hartree-Fock approximation (HF); \textit{red points:} result from the DMRG for a chain with $36$ cells.}
 \end{figure}
\begin{figure}[t]
 \includegraphics[width=85mm]{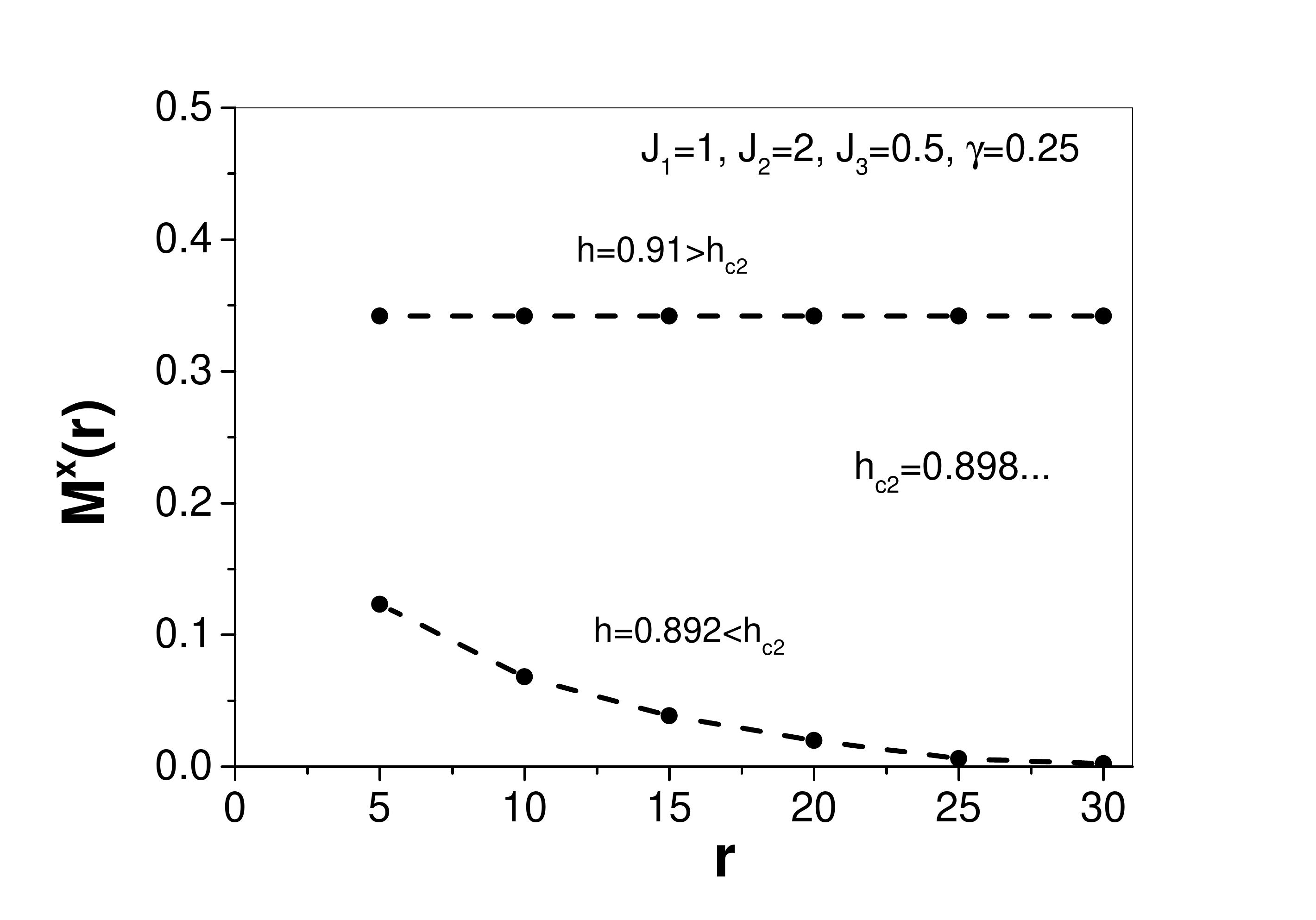}%
 \caption{\label{fig_assitota_temp}Asymptotic behaviour of $M^{x}(r)=\sqrt{\langle S_{j}^{x}S_{j+r}^{x}  \rangle}$ as function of $r$ for the distorted diamond chain with $J_{1}=1$, $J_{2}=2$, $J_{3}=0.5$ and $\gamma=0.25$ , in the region of the critical point $h_{c2}=0.898... $, for $h=0.892$ and $h=0.91$  }
 \end{figure}

The Fig,\ref{fig_assitota_temp} shows the asymptotic behaviour for  $M^{x}(r)=\sqrt{\langle S_{j}^{x}S_{j+r}^{x}  \rangle}$ as a function of $r$ for the distorted chain with $J_{1}=1$, $J_{2}=2$, $J_{3}=0.5$, $\gamma=0.25$, for two values of field $h=0.892$ and $h=0.91$, in the region of the critical point  $h_{c2}=0.898... $, where we expect quantum fluctuations to be greatest. These results confirm that $M^{x}(r=30)\cong M^{x}$, which justifies the fact that we used $r=30$ cells in Figs. \ref{fig_mx_susc_j1_1_j2_1_j3_1_g_025} and \ref{fig_mx_susc_j1_1_j2_2_j3_05_g_025}.
 
 \begin{figure}[t]
 \includegraphics[width=85mm]{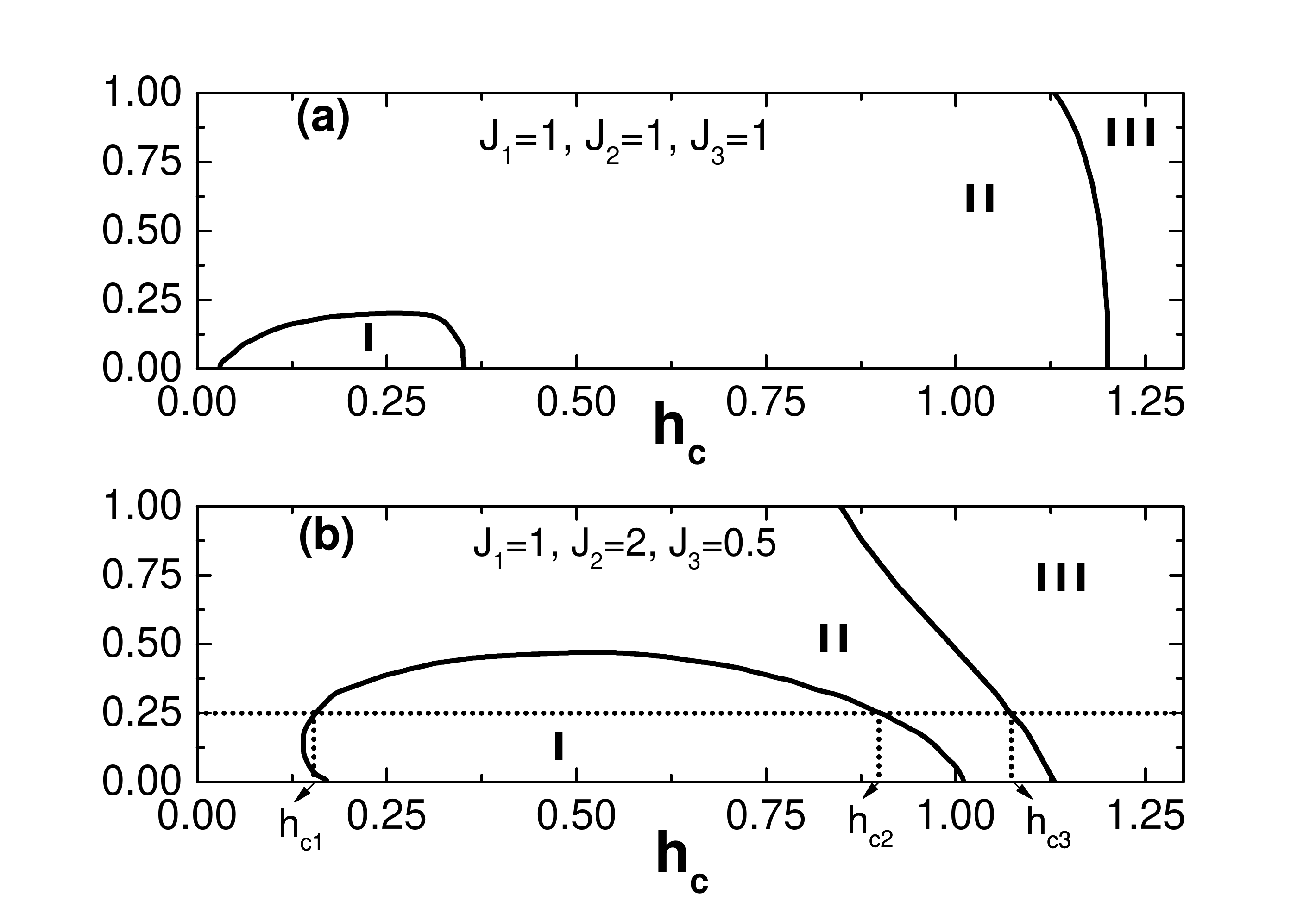}%
 \caption{\label{fig_hc}Typical critical lines at $T=0$ for a diamond chain: \textbf{(a)} for a symmetrical configuration $J_{1}=J_{2}=J_{3}=1$, $\gamma=0.25$, there is only one critical point; \textbf{(b)} for a distorted configuration, $J_{1}=1$, $J_{2}=2$, $J_{3}=0.5$ and $\gamma =0.25$. For $\gamma=0.25$ we identify the three critical points, $h_{c1}=0.161...$, $h_{c2}=0.898...$, $h_{c3}=1.059...$}
 \end{figure}

These results show that the HF approximation handles the calculation of critical points for the XY model, apart from slight changes in its precise values.
Thus, as our key result, it enables us to repeat the same procedure to other values of $\gamma$ (within this mean-field approach), and obtain the phase diagrams $\gamma\times h_{c}$ presented in Fig.\,\ref{fig_hc} for (a) $J_{1}=1$, $J_{2}=2$, $J_{3}=0.5$, and (b) $J_{1}=J_{2}=J_{3}=1$.
Here, region I, II, and III correspond to the gapped, gapless, and saturated phases, respectively.
Notice that, for small values of $\gamma$, one always has three critical points, while when it increases, just a single critical point remains in the phase diagrams.
This feature is similar to those of the XY model over a linear periodic chain of period three, as can be seen in \citep{lima2007}.
Furthermore, as seen in Fig.\,\ref{fig_hc}, the 1/3-plateau phase is way reduced in the symmetrical chain, as an effect of the strong frustration in this case.

\begin{figure}[t]
 \includegraphics[width=85mm]{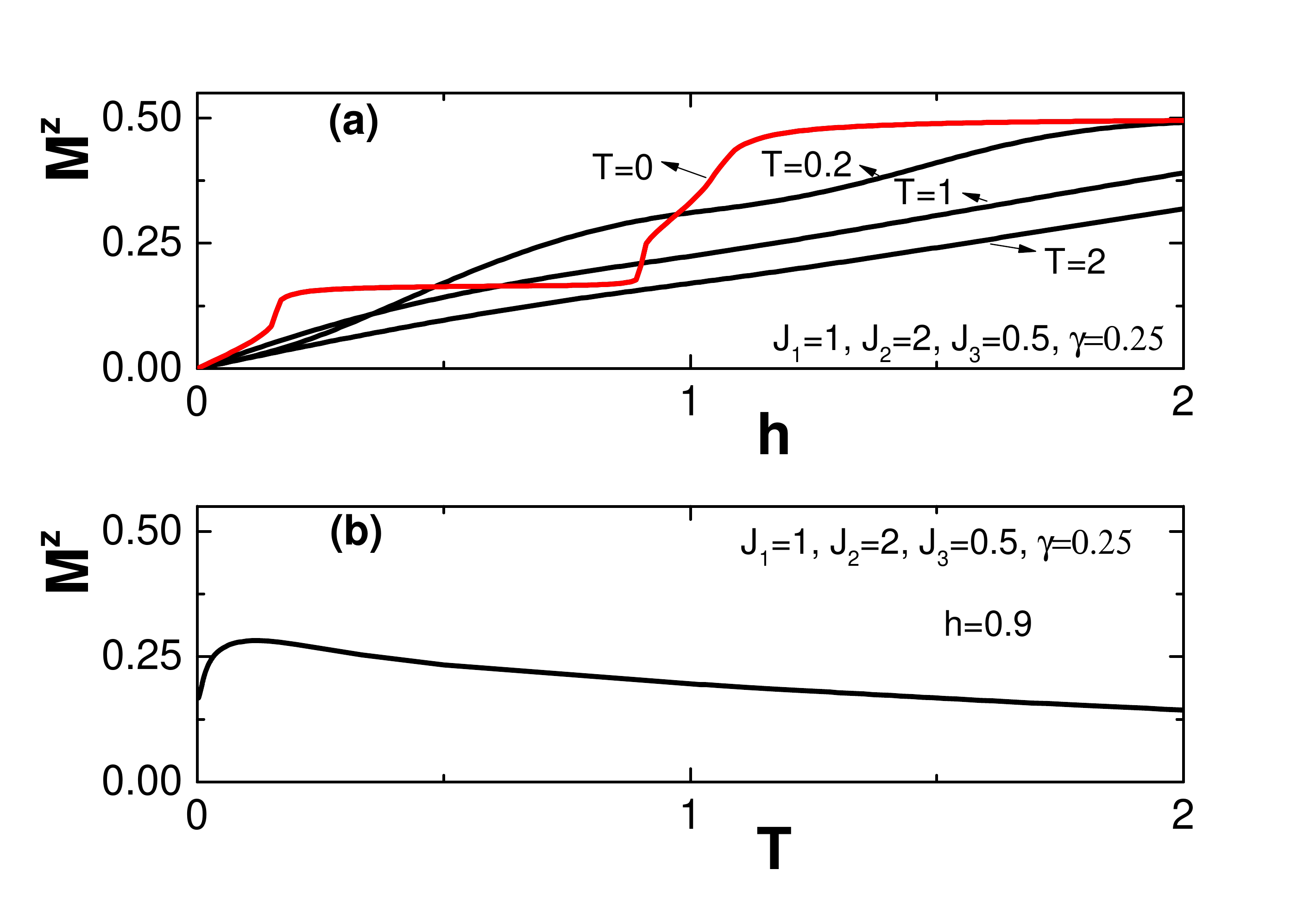}%
 \caption{\label{fig_mag_z_temp_diff_zero} \textbf{(a)} Induced magnetization as a function of the field, for a distorted diamond chain  $J_{1}=1$, $J_{2}=2$, $J_{3}=0.5$ and $\gamma=0.25$ for temperature $T=0,\ 0.2,\ 1,\ 2$. Here, was made $K_{B}=1$. \textbf{(b)} Induced magnetization as a function of temperature, for a distorted diamond chain with $J_{1}=1$, $J_{2}=2$, $J_{3}=0.5$, $\gamma=0.25$ and $h=0.9$.   }
 \end{figure}

An advantage of the HF over ED or DMRG methods is that it allows us to study the effects of temperature.
Figure \ref{fig_mag_z_temp_diff_zero} shows $M^{z}\times h$ for a distorted chain with $J_{1}=1$, $J_{2}=2$, $J_{3}=0.5$ and $\gamma=0.25$, at different temperature scales.
Simililarly to $\gamma$, the temperature also destroys the plateaus, but due to thermal fluctuations.
Indeed, the intersection points on the curves arise from the difference in intensity between thermal and quantum fluctuations.

Interestingly, for some range of $h$, we find an unusual behavior for the induced magnetization.
For instance, fixing $h=0.9$, and setting the same parameters of the Fig.\,\ref{fig_mag_z_temp_diff_zero}\,(a), $M^{z}$ increases with temperature, as exhibited in Fig.\,\ref{fig_mag_z_temp_diff_zero}\,(b).
Indeed, this curve has a hump that agrees with the intersection points in the curves of the Fig.\,\ref{fig_mag_z_temp_diff_zero}\,(a), for $h$ values away from the saturation region.
In the region where the curve is ascending the quantum ordering prevails driven by the predominance of quantum fluctuations, and in the region where $M^{z}$ decreases with $T$, the classical fluctuations predominate, just like in classical/quantum crossovers, as discussed in Refs.\,\cite{young2001,goncalves2005}.
However, this issue is beyond of the scope of this work, and further investigations are on-going.
 
 \section{Conclusions}
 \label{conclusions}
In this paper we consider the XY model on a diamond chain for symmetrical and distorted cases. The Hamiltonian was fermionized by Jordan-Wigner transformations, and the quartic terms were treated by a Hartree-Fock approximation.
Given this, we determined approximate expressions for induced magnetization, isothermal susceptibility, static correlation functions $\langle S_{j}^{x}S_{j+r}^{x}\rangle$, and spontaneous magnetization at any temperature.
The results obtained by the mean-field approach were compared with ED and DMRG calculations, for chains with $N=8$ and 36 unit cells, respectively.
Interestingly, we noticed that the HF approximation works better as the anisotropy parameter increases.
Indeed, we have found a good agreement between the mean-field and these unbiased methodologies concerning the magnetic phases and critical points.

Through the analysis of singularities in the isothermal susceptibility, it was possible to obtain phase diagrams $\gamma\times h$ at the ground state.
Then, we could understand the behavior of the phases (gapped, gapless, or saturated) under changes in the external parameters.
In particular, the region of the 1/3-plateau phase is very reduced when $J_{1}=J_{2}=J_{3}$, and is suppressed when $\gamma$ increases.
At finite temperatures, we found an unusual behavior for the induced magnetization away from the saturation region, from which $M^{z}$ increases with temperature -- a feature related to classical/quantum crossovers.
In summary, this work presents a \textit{global} picture of the quantum XY model on the frustrated diamond chain, which may be useful to understand features of magnetism in more complex frustrated geometries. 



\section*{Acknowledgements}
The authors are grateful to R.R.\,dos Santos for discussions, and to the Brazilian Agencies CAPES, CNPq, FAPEPI, and
FAPERJ for financial support.

\bibliographystyle{elsarticle-num}
\bibliography{bib_diamond_chain}


\end{document}